# A Methodology to Generate Virtual Patient Repositories

Uri Kartoun PhD, Massachusetts General Hospital / Harvard Medical School

## ABSTRACT

Electronic medical records (EMR) contain sensitive personal information. For example, they may include details about infectious diseases, such as human immunodeficiency virus (HIV), or they may contain information about a mental illness. They may also contain other sensitive information such as medical details related to fertility treatments. Because EMRs are subject to confidentiality requirements, accessing and analyzing EMR databases is a privilege given to only a small number of individuals. Individuals who work at institutions that do not have access to EMR systems have no opportunity to gain hands-on experience with this valuable resource. Simulated medical databases are currently available; however, they are difficult to configure and are limited in their resemblance to real clinical databases. Generating highly accessible repositories of virtual patient EMRs while relying only minimally on real patient data is expected to serve as a valuable resource to a broader audience of medical personnel, including those who reside in underdeveloped countries.

## INTRODUCTION

The importance of patient privacy has been thoroughly emphasized by governmental resources such as the HIPAA Privacy Rule and by academia [1–3]. Numerous strategies to maintain patients' privacy have been developed [4–8]; however, the medical profession is not yet able to guarantee full protection of privacy while providing detailed information about each patient [9].

Cohorts assembled from electronic medical records (EMRs) represent a powerful resource to study disease complications at a population level. Recent studies have demonstrated the usefulness of EMR analysis for discovering or confirming outcome correlations, subcategories of disease, and adverse drug events [10–16]. Due to confidentiality restrictions, accessing and analyzing EMR databases is a privilege given to only a small number of individuals. Individuals who work in institutions that do not have access to EMR systems cannot experiment with such valuable resources. When professors wish to teach a biomedical informatics course focused on EMR technology, they cannot distribute real EMR data among their students.

Simulated medical databases [e.g., 17, 18] and open EMR platforms [19] are currently available; however, no confidentiality-free massive scale longitudinal EMR databases have yet been algorithmically created. Virtual patient repositories that bear a high degree of resemblance to real patient databases while relying only minimally on real patient data are expected to serve as a valuable resource for medical professionals in training, and to accelerate health care research and development.

The aim of this study is to develop a novel methodology for creating virtual patient repositories. I demonstrate that a method entailing minimal configuration requirements can generate nonconfidential artificial EMR databases that could be used to practice statistical and machine-learning algorithms. I further demonstrate the potential broad public interest in the availability of this technology.

## MATERIALS AND METHODS

The process of generating a virtual patient repository was based on preconfiguration of population-level and patient-level characteristics. First, an object-oriented program acquired a population-level configuration to generate patient objects. Next, the program created a clinical profile for each patient, including admissions associated with chief complaints and laboratory measurements. Finally, the patient objects were stored in a database. Following this methodology, three databases of 100, 10,000, and 100,000 virtual patients were created.

### Population-level configuration

Population-level configuration specifies the number of individual records that will be generated and defines preconfigured values for demographic characteristics. Demographic characteristics include gender, marital status, major language, ethnicity, date of birth, and income level. Configuring categorical variables defines several potential values for the variable and the percentage of the population with the value. For example, in a population of $n = 100,000$ individuals, a potential configuration for ethnicity would be 49% white, 23% Asian, 15% African American, and the remainder unknown. For continuous variables such as age, the population percentages for several ranges of date of birth are defined. For example, dates of birth in the range of 1940 to 1950 were randomly created for 15% of the population. The configuration tables are presented in Supplemental Table 1(a–b).

### Generating a virtual patient repository

Having acquired the population-level configuration, the program generated $n$ objects each representing one virtual patient. Each such object is associated with distinct demographic characteristics (e.g., a certain patient might be an English-speaking, Asian, single woman who was born on May 17, 1982). For each such object, the program randomly assigns an integer representing the number of admissions associated with that patient throughout his or her lifetime. Each admission is associated with additional details, including randomly generated length of stay (in days) and start and end dates. Further, each admission is associated with laboratory measurements and a chief complaint randomly selected from a list of International Classification of Diseases, Tenth Revision, Clinical Modification (ICD-10-CM) codes. Laboratory values are based on 35 common types—for instance, sodium levels, creatinine levels, or platelet count. The program generates for each admission multiple dates and times for each laboratory type; values are bounded by predefined ranges for each laboratory type. For example, a patient may be associated with a 3-day admission during which creatinine was measured 6 times throughout the admission. The configuration table for laboratory types and ranges is presented in Supplemental Table 2. An example for the EMR lifespan of one virtual patient is presented in Figure 1.

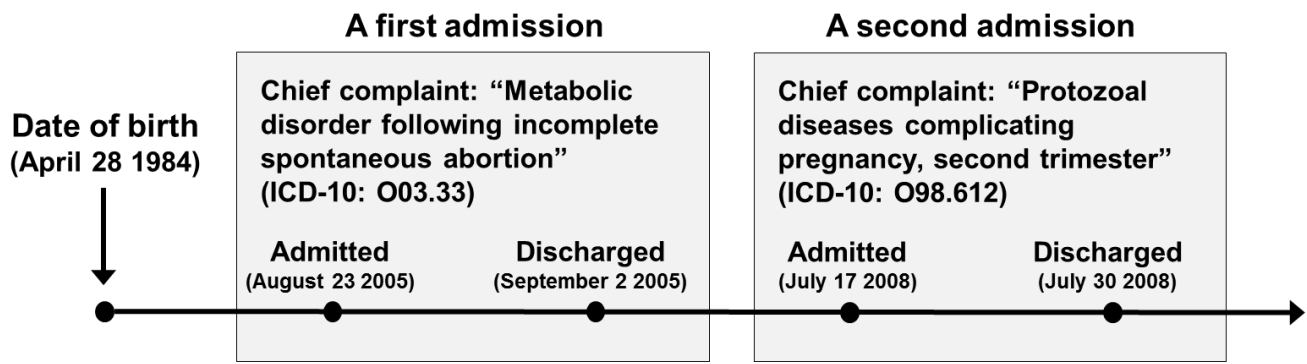

Figure 1. An example of the longitudinal EMR of a virtual patient.

## RESULTS

### A virtual patient repository

Table 1 summarizes several characteristics of the largest cohort, which comprised 100,000 virtual patients. Each virtual patient in the cohort is associated with 1 to 10 admissions; each admission lasts from 1 to 20 days and is associated with a single chief complaint. All values are associated with a date and time. Each admission record also includes multiple measurements of common laboratory tests (see Supplemental Table 2). The number of admissions per patient, the length of stay per admission, and the laboratory values were randomly generated; however, they are sampled from predefined ranges of values. For example, a patient's age could not exceed 95 years old as of January 1, 2015, and a glucose measurement could only be in the range of 60–140 mg/dL. Only chief complaints common for both men and women were allowed in a cohort. In total, the database contained

1.4 GB of data, representing 100,000 virtual patients associated with 361,760 admissions and 107,535,387 total laboratory measurements.

**Table 1. A virtual patient repository of 100,000 patients.**

| Variable and category | Patients (n = 100,000) |
|---|---|
| **Mean age as of 1/1/2015, years (SD)** | 57.8 (17.3) |
| **Gender (%)** | |
| Female | 52.0 |
| **Ethnicity (%)** | |
| White | 49.0 |
| Asian | 23.0 |
| African American | 15.0 |
| Unknown | 13.0 |
| **Mean number of admissions per patient, days (SD)** | 3.6 (1.5) |
| **Mean length of stay (SD)** | 11.0 (5.2) |
| **% Population with length of follow-up (years)** | |
| 0 - 9 | 13.1 |
| 10 - 15 | 9.3 |
| > 15 | 77.6 |
| **Population below poverty (%)** | 21.6 |
| **Comorbidities; Prevalence (%)** | |
| Malignant neoplasm | 41.4 |
| Rheumatoid arthritis | 25.6 |
| Diabetes (type I or II) | 24.4 |
| Renal complications | 17.0 |
| Coronary artery disease | 7.0 |
| **Laboratory values (Mean; SD)** | |
| Blood urea nitrogen (mg/dL) | 17.5; 7.2 |
| Platelets (k/cumm) | 284.9; 95.3 |
| Creatinine (mg/dL) | 0.9; 0.2 |
| Albumin (gm/dL) | 4.2; 1.0 |
| Lymphocytes (%) | 25; 5.8 |

## DISCUSSION

In April 2015, the three generated cohorts were made publically available for downloading at the dedicated website www.emrbots.org. Since its launch, the site has been visited by more than 5,000 individuals; approximately 1,000 visitors registered with their full names, institution names, and e-mail addresses. The represented institutions included top American universities, pharmaceutical companies, and governmental agencies such as the Centers for Disease Control and Prevention (CDC).

The virtual patient repositories can be instantaneously downloaded to allow users to practice algorithms on EMR-like data with a primary objective of improving their technical skills. These skills may be used in the future when the individuals are granted access to real EMR. Using the repositories would not require installation of any software. The repositories contain raw data only and are provided as textual flat files; thus they are independent of any specific operating system, and may be used with a wide variety of database systems. Further, the repositories can be used even on low-performance computers, as they can be run on open-source software products—such as R and MySQL—which have minimal requirements to install. The repositories could also be deployed on existing open-source EMR platforms [19]. With no confidentiality restrictions, the cohorts can be distributed to a class of any size. For example, professors could distribute the records to a class of 100 students, all of whom could then use them to practice algorithms and to build models; at the end of the course, these students would have greater experience and capabilities, and would thus be more attractive to employers in the field of medical informatics.

While the cohorts are useful for practicing machine-learning algorithms, they cannot serve as effective resources to assess real patient outcome scenarios (e.g., 30-day readmission prediction or disease prognosis) because their creation process does not take into account the complex time-dependent interactions between the factors associated with real patients. Developing algorithms to create virtual patient repositories that will reliably mimic real EMR presents is a tremendous challenge because of it necessitates approaches that will populate databases with a combination of linear and nonlinear associations between all medical elements, as well as with random associations. The algorithms required would comprise both individual-level and population-level assumptions and apply an intelligent functionality to assign acceptable temporal differences among all medical events. A virtual patient repository of nonalcoholic fatty liver disease patients, for example, would need to include assumptions about inverse correlations of albumin levels and sodium levels with cardiovascular disease [20], while virtual congestive heart failure (CHF) patients would need to be associated with a high prevalence of diuretic use, advanced age, and a high prevalence of associated comorbidities, such as renal failure [21].

A fundamental requirement is to define appropriate performance measures to assess the quality of the generated virtual data. Achieving a high level of resemblance between virtual and real patient data will necessitate development of methods to identify similarities among patients. Such methods are available; however, they focus either on several predefined variables [e.g., 22] based on analyzing images [e.g., 23] or on genetic similarities in patients [e.g., 24]. There are as yet no available methods that consider all the complex relationships among EMR variables. Current strategies for identifying groups of patients with similar time-dependent characteristics include time-series analysis methods [e.g., 25–27] and clustering techniques [e.g., 11, 28–32]. Such methods are based on either the supervised learning paradigm (which requires extracting manually designed features) or on the unsupervised learning paradigm (in which candidate variables are computationally proposed). The task of identifying patients with similar characteristics is multidimensional. For example, two patients might have similar diagnoses (e.g., both had a recent heart failure episode, or both are diabetic); however, they might vary in many characteristics (e.g., comorbidities or laboratory measurements). Identifying distinct groups of patients with similar laboratory trends, correlated with comorbidities and medications changing over time and with demographics, is expected to accelerate the development of algorithms that will more realistically create virtual patients.

In conclusion, this study describes a novel methodology for generating confidentiality-free virtual patient repositories. The methodology may serve as a foundation to further generate large, longitudinal artificial EMR databases that highly resemble real patient records. Unlike other methods that typically obscure or shift real patients' data elements, the proposed methodology is invulnerable in terms of security because it does not rely on real data elements pulled from an existing EMR; therefore, it is not associated with privacy concerns in regard to individuals' sensitive data.

## REFERENCES

1. Barrows RC Jr, Clayton PD. Privacy, confidentiality, and electronic medical records. J Am Med Inform Assoc 1996;3(2):139–48.

2. Lamberg L. Confidentiality and privacy of electronic medical records: psychiatrists explore risks of the "information age". J Am Med Inform Assoc 2001;285(24):3075–6.

3. Caine K, Hanania R. Patients want granular privacy control over health information in electronic medical records. J Am Med Inform Assoc 2013;20(1):7–15.

4. Benaloh J, Chase M, Horvitz E, et al. Patient controlled encryption: ensuring privacy of electronic medical records. Proceedings of the 2009 ACM workshop on Cloud computing security. pp. 103–14.

5. Hampton KH, Fitch MK, Allshouse WB, et al. Mapping health data: improved privacy protection with donut method geo-masking. Am J Epidemiol 2010;172(9):1062–9.

6. Pal D, Chen T, Zhong S, et al. Designing an algorithm to preserve privacy for medical record linkage with error-prone data. J Med Internet Res 2014;2(1):e2.

7. Heatherly R, Rasmussen LV, Peissig PL, et al. A multi-institution evaluation of clinical profile anonymization. J Am Med Inform Assoc 2016;23(e1):e131–7.

8. Hripcsak G, Mirhaji P, Low AF, et al. Preserving temporal relations in clinical data while maintaining privacy. J Am Med Inform Assoc 2016;pii:ocw001.

9. Kohane IS, Drazen JM, Campion EW. A glimpse of the next 100 years in medicine. N Engl J Med 2012;367:2538–9.

10. Liao KP, Kurreeman F, Li G, et al. Associations of autoantibodies, autoimmune risk alleles, and clinical diagnoses from the electronic medical records in rheumatoid arthritis cases and non-rheumatoid arthritis controls. Arthritis Rheum 2013;65(3):571–81.

11. Doshi-Velez F, Ge Y, Kohane I. Comorbidity clusters in autism spectrum disorders: an electronic health record time-series analysis. Pediatrics 2014;133(1):e54–e63.

12. Liao KP, Cai T, Savova GK, et al. Development of phenotype algorithms using electronic medical records and incorporating natural language processing. BMJ 2015;350:h1885.

13. Wilke RA, Xu H, Denny JC, et al. The emerging role of electronic medical records in pharmacogenomics. Clin Pharmacol Ther 2011;89(3):379–86.

14. Roden DM, Xu H, Denny JC, et al. Electronic medical records as a tool in clinical pharmacology: opportunities and challenges. Clin Pharmacol Ther 2012;91(6):1083–86.

15. Denny JC, Bastarache L, Ritchie MD, et al. Systematic comparison of phenome-wide association study of electronic medical record data and genome-wide association study data. Nat Biotechnol 2013; 31(12):1102–10.

16. Bowton E, Field JR, Wang S, et al. Biobanks and electronic medical records: enabling cost-effective research. Sci Transl Med 2014;6(234).

17. Buczak AL, Babin S, Moniz L. Data-driven approach for creating synthetic electronic medical records. BMC Med Inform Decis Mak 2010;10:59.

18. Moniz L, Buczak AL, Hung L, et al. Construction and validation of synthetic electronic medical records. Online J Public Health Inform 2009;1(1):ojphi.v1i1.2720.

19. Zaidan AA, Zaidan BB, Al-Haiqi A, et al. Evaluation and selection of open-source EMR software packages based on integrated AHP and TOPSIS. J Biomed Inform 2015;53:390–404.

20. Corey KE, Kartoun U, Zheng, et al. Electronic medical records database to identify non-traditional cardiovascular risk factors in nonalcoholic fatty liver disease. Am J Gastroent 2016;111(5):671–6.

21. Allen LA, Smoyer Tomic KE, Smith DM, et al. Rates and predictors of 30-day re-admission among commercially insured and Medicaid-enrolled patients hospitalized with systolic heart failure. Circ Heart Fail 2012;5(6):672–9.

22. Gottlieb A, Stein GY, Ruppin E, et al. A method for inferring medical diagnoses from patient similarities. BMC Medicine 2013;11:194.

23. Buyue Q, Xiang W, Nan C, et al. A relative similarity based method for interactive patient risk prediction. Data Min Knowl Discov 2015;29(4):1070–93.

24. Wang B, Mezlini AM, Demir F, et al. Similarity network fusion for aggregating data types on a genomic scale. Nat Methods 2014;11(3):333–7.

25. Dowding DW, Turley M, Garrido T. The impact of an electronic health record on nurse sensitive patient outcomes: an interrupted time series analysis. J Am Med Inform Assoc 2012;19(4):615–20.

26. Rathlev NK, Obendorfer D, White LF, et al. Time series analysis of emergency department length of stay per 8-hour shift. West J Emerg Med 2012;13(2):163–8.

27. Duerichen, R, Pimentel MAF, Clifton L, et al. Multi-task Gaussian processes for multivariate physiological time-series analysis. IEEE Trans Biomed Eng 2015;62(1):314–22.

28. Shah SJ, Katz DH, Selvaraj S, et al. Phenomapping for novel classification of heart failure with preserved ejection fraction. Circulation 2015;131(3):269–79.

29. Li L, Cheng WY, Glicksberg BS, et al. Identification of type 2 diabetes subgroups through topological analysis of patient similarity. Sci Transl Med 2015;7(311):311ra174.

30. Ng K, Ghoting A, Steinhubl SR, et al. PARAMO: a PARAllel predictive MOdeling platform for healthcare analytic research using electronic health records. J Biomed Inform 2014;48:160–70.

31. Gillam MT, Cazangi RR, Kontsevoy AP, et al. Automated clustering for patient disposition. Publication number: US8589187. 2013.

32. Fabbri D, Lefevre K. Explaining accesses to electronic medical records using diagnosis information. J Am Med Inform Assoc 2013;20(1):52–60.

**Supplemental Table 1. Population-level configuration**

**(a) Categorical variables**

| Variable | Value | Weight (%) |
|---|---|---|
| Gender | Male | 48 |
| | Female | 52 |
| Marital status | Single | 32 |
| | Married | 45 |
| | Divorced | 11 |
| | Separated | 5 |
| | Widowed | 1 |
| | Unknown | 6 |
| Language | English | 64 |
| | Icelandic | 12 |
| | Spanish | 18 |
| | Unknown | 6 |
| Ethnicity | African American | 15 |
| | Asian | 23 |
| | White | 49 |
| | Unknown | 13 |

**(b) Continuous variables**

| Variable | Range Min | Range Max | Weight (%) |
|---|---|---|---|
| Date of birth | 1/1/1920 | 1/1/1930 | 8 |
| | 1/1/1930 | 1/1/1940 | 9 |
| | 1/1/1940 | 1/1/1950 | 15 |
| | 1/1/1950 | 1/1/1960 | 19 |
| | 1/1/1960 | 1/1/1970 | 24 |
| | 1/1/1970 | 1/1/1980 | 15 |
| | 1/1/1980 | 1/1/1990 | 10 |
| Population percentage below poverty | 0 | 10 | 10 |
| | 10.1 | 20 | 80 |
| | 80.1 | 100 | 10 |

**Supplemental Table 2. Laboratory type and value configuration**

| Laboratory title | Minimal allowed value | Maximal allowed value | Units |
|---|---|---|---|
| White blood cell count | 3 | 12 | k/cumm |
| Red blood cell count | 3 | 7 | m/cumm |
| Hemoglobin | 10 | 19 | gm/dl |
| Hematocrit | 30 | 55 | % |
| Mean corpuscular volume | 70 | 100 | fl |
| Mean cell hemoglobin | 22 | 40 | pg |
| Mean corpuscular hemoglobin concentration | 28 | 40 | g/dl |
| Red cell distribution width | 9 | 17 | % |
| Platelet count | 120 | 450 | k/cumm |
| Absolute neutrophils | 60 | 80 | % |
| Absolute Lymphocytes | 15 | 35 | % |
| Neutrophils | 1 | 11 | k/cumm |
| Lymphocytes | 0.5 | 5 | k/cumm |
| Monocytes | 0.08 | 1.2 | k/cumm |
| Eosinophils | 0.08 | 0.6 | k/cumm |
| Basophils | 0 | 0.22 | k/cumm |
| Sodium | 125 | 155 | mmol/L |
| Potassium | 3 | 6 | mmol/L |
| Chloride | 90 | 115 | mmol/L |
| Carbon dioxide | 18 | 36 | mmol/L |
| Anion gap | 3 | 18 | mmol/L |
| Glucose | 60 | 140 | mg/dL |
| Blood urea nitrogen | 5 | 30 | mg/dL |
| Creatinine | 0.5 | 1.2 | mg/dL |
| Total protein | 5 | 10 | gm/dL |
| Albumin | 2.5 | 6 | gm/dL |
| Calcium | 7 | 12 | mg/dL |
| Total bilirubin | 0 | 1.2 | mg/dL |
| Aspartate aminotransferase | 12 | 42 | U/L |
| Serum glutamic pyruvic transaminase | 15 | 75 | U/L |
| Alkaline phosphatase | 40 | 150 | U/L |
| Urine specific gravity | 1.014 | 1.028 | no unit |
| Urine pH | 4.5 | 7.5 | no unit |
| Urine red blood cells | 0 | 3.5 | rbc/hpf |
| Urine white blood cells | 0 | 6 | wbc/hpf |